\documentclass[12pt,preprint]{aastex}
\usepackage{emulateapj5}

\newcommand{\etal}{et~al.}
\newcommand{\ie}{{\it i.e.}}
\newcommand{\eg}{{\it e.g.}}
\newcommand{\cf}{{\rm cf.}}
\newcommand{\teff}{$T_{\rm eff}$}
\newcommand{\msun}{M$_{\sun}$}
\newcommand{\ic}{$I_{\rm C}$}

\newcommand{\mdot}{$\dot{M}$}

\received{}
\revised{}
\accepted{}
\shorttitle{Stellar Rotation and Radius}
\shortauthors{Rebull \etal}

\begin{document}

\title{On the Relationship between Stellar Rotation and Radius in 
Young Clusters}
\author{L.\ M.\ Rebull\altaffilmark{1}, S.\ C.\ Wolff\altaffilmark{2},   
	S.\ E.\ Strom\altaffilmark{2}, R.\ B.\ Makidon\altaffilmark{3}}

\altaffiltext{1}{National Research Council Resident Research Associate,
        NASA/Jet Propulsion Laboratory, M/S 169-506, 4800 
        Oak Grove Drive, Pasadena, CA 91109 (luisa.rebull@jpl.nasa.gov)}
\altaffiltext{2}{NOAO, 950 N.\ Cherry Ave, Tucson, AZ 87526}
\altaffiltext{3}{STScI, 3700 San Martin Dr, Baltimore, MD 21218}

\begin{abstract}

We examine the early angular momentum history of stars in young clusters
via 197 photometric periods in the Orion Flanking Fields, 83 photometric
periods in NGC 2264, and 256 measurements of $v \sin i$ in the ONC.  We
show that PMS stars, even those without observable disks, apparently do
not conserve stellar angular momentum as they evolve down their convective
tracks, but instead evolve at nearly constant angular velocity.  This
result is inconsistent with expectations that convective stars lacking
disks should spin up as they contract, but paradoxically consistent with
disk-locking models.  We briefly explore possible resolutions, including
disk locking, birthline effects, stellar winds, and planetary companions. 
We have found no plausible explanations for this paradox.

\end{abstract}


\section{Introduction}

Observations over the past decade have established that most--and probably
all--pre-main sequence (PMS) stars are surrounded by disks during early
stages of their evolution, and that a substantial portion of their final
mass is accreted from these disks.  However, the prediction that this
accretion will cause the stars to rotate at close to their breakup speed
(\eg\ Durisen \etal\ 1989) is not borne out by observations.  Most PMS
stars have rotational velocities ($v$) of no more than a few tens of km
s$^{-1}$.  By contrast, the typical breakup velocity at $\sim$1 Myr is
$\sim$300 km s$^{-1}$.

Explanations of the observed slow rotation have invoked a magnetic field
that is rooted in the central star and intercepts the disk (K\"onigl
1991).  The net consequence of this interaction is a braking torque
transmitted to the star by field lines penetrating the disk beyond the
co-rotation radius (where the angular velocity $\omega$ of the disk
matches that of the star).  This braking torque exactly balances the
spin-up torque exerted by the material inside the co-rotation radius. 
This model predicts that stars will be locked to their disks and will
rotate slowly until the time that the disk is dissipated.  Other models
(\eg\ Shu \etal\ 2000) make use of a wind, launched from the stellar
magnetosphere-disk boundary to carry away stellar angular momentum ($J$),
thus reducing rotation rates to well below breakup.  By contrast, stars
without disks should conserve $J$ and spin up as they contract and evolve
down their Hayashi tracks. 

Considerable observational effort has been devoted to finding empirical
correlations between rotation rates and the presence or absence of
observable disks; although such correlations appeared in early studies
(\eg\ Edwards \etal\ 1993, Choi \& Herbst 1996), recent results are 
ambiguous at best (\eg\ Rebull 2001; Stassun \etal\ 1999). Over the past
few years, rotation periods $P$ or projected rotational velocities $v \sin
i$ have been measured for large samples of PMS stars $\sim$0.1-2.5 \msun\
and of age $\lesssim$3 Myr.  These data sets are now large enough to
provide direct evidence of how $J$ changes as stars of different masses
evolve down their convective tracks.  In this paper, we examine the early
$J$ history of stars surrounded by disks and those that lack disks.  We
show that PMS stars, {\it even those without observable disks}, do not
conserve angular momentum as they evolve toward the ZAMS but instead
evolve at nearly constant angular velocity.  This result is inconsistent
with expectations that convective stars lacking disks should spin up as
they contract, but paradoxically consistent with disk-locking models.  

\section{The Observations}

The data sets that we draw on for this paper are 256 measurements of $v
\sin i$ in the Orion Nebula Cluster (ONC) by Rhode \etal\ (2001; RHM); 197
photometric periods ($P$) for the Flanking Fields (FF) that surround the
ONC (Rebull 2001, Rebull \etal\ 2000); and 83 stars with $P$ in NGC 2264
(Makidon \etal\ 2001, Rebull \etal\ 2001).  The age distribution for Orion
peaks near $\sim$1 Myr, while NGC 2264 peaks near $\sim$1 Myr but contains
significant numbers of stars out to $\sim$3-4 Myr.  Together, these
clusters allow us to examine changes in stellar angular momenta over a
total range in stellar radius of $\sim$0.4 dex. In order to estimate
accurate stellar radii and to determine the presence or absence of a disk,
we required that spectral types along with $I-K$ and $V-I$ photometry be
available for all the stars included in the present study. While other
disk indicators (\eg\ UV excess, H$\alpha$ emission) are available for
subsets of the Orion and NGC 2264 databases, only $I-K$ is widely
available.  All stars with an $I-K$ color redder by 0.3 mag than expected
based on spectral type were considered to be surrounded by disks (\cf\
Rebull 2001 and references therein).  We note that 26/197 (13\%) stars in
the Orion FF sample and 20/83 (24\%) in the NGC 2264 sample have $I-K$
excesses, whereas 60\% of the stars in the ONC sample have such an
excess.  

Stellar temperatures (\teff) were assigned according to spectral types and
luminosities ($L$) were derived from \ic\ magnitudes with reddening 
estimated from colors and spectral types following Hillenbrand (1997).  We
adopted distance moduli of 8.36 for Orion (Genzel \etal\ 1981) and 9.40
for NGC 2264 (Sung \etal\ 1997).  Radii ($R$) were calculated using the
assigned values of $L$ and \teff.  Photometric uncertainty, the amplitude
of the periodic modulation, uncertainties in the reddening correction, and
errors in classification all contribute to errors in $L$ and \teff\ and
hence to uncertainty in $R$.  The distribution of empirically-determined
errors in $\log R$ is strongly peaked near $\pm$0.1.  The errors in $P$
are less than 1\%.

The $v \sin i$ data for the ONC have sufficient resolution to measure only
velocities greater than $\sim$11 km s$^{-1}$, corresponding to $P\lesssim
6$d for a typical PMS star in our sample.  The $P$ measurements are
sensitive to a much broader range of periods, viz $P\sim0.3-25$d ($v\sim
3-300$ km s$^{-1}$), encompassing much slower rotation velocities than do 
the published $v \sin i$ measurements.  Approximately half of the stars
with $v \sin i$ measurements by RHM were chosen because they had known
$P$.  The other half were selected as a control sample of objects without
known $P$ from the same portion of the HR diagram.  RHM show there is no
statistically significant difference between the $v \sin i$ distributions
of the periodic and control samples, leading to the important conclusion
that samples of stars found to show photometric periods can be taken as a
fair representation of the range of rotational characteristics of PMS
stars.

\section{Analysis}

If stars conserve angular momentum ($J$) as they evolve down their
convective tracks, then we would expect that $J=MvR$ would remain
constant, but only if $J$ is conserved in the outermost observable layer
of the star.  We expect fully convective stars to rotate as solid bodies. 
Examination of theoretical models of PMS stars (\eg\ Swenson \etal\ 1994)
indicates that the predicted changes in surface rotation rate as the star
evolves down the convective track are indistinguishable for the two
extreme cases of solid body rotation and local conservation of $J$ (\ie\
$J$ conservation in spherical shells).  Along the convective track
contraction is nearly homologous, and changes in $I$ directly track
changes in $R^2$.  If $J$ is conserved, then $vR$ is constant and $P =
2\pi R/v$ should be $\propto R^2$.  On the other hand, if angular velocity
$\omega = v/R$ is constant, as predicted for disk-locking, then $P$ is
constant.

In Figure 1, we plot $P$ and $j=\frac{2Rv}{5}$ vs.\ $R$, where $j$ is
computed as in Rebull (2001)\footnote{Computed assuming spherical stars
rotating as solid bodies, $j=\frac{J}{M}=\frac{I\omega}{M}=\frac{2Rv}{5} =
\frac{4\pi R^2}{5P}$, where $I$=moment of inertia, and all other terms
have been defined.}.  It is immediately apparent that there is no
variation of $P$ with $R$, but that $j$ changes with $R$, consistent with
conservation of $\omega$.  The errors in log $R$ would have to be about
0.5 dex (5 times larger than we estimate them to be) in order to mask the
decline in $P$ expected for the case of conservation of $J$.  We see no
substantial difference between these two clusters, except for the fact
that the stars in NGC 2264 include older objects with smaller radii.  We
find no significant difference in the period-radius relationship of stars
with and without $I-K$ excesses, though there may be some marginal
tendency for diskless stars to have the shortest periods. We have divided
our sample into two bins: $P<$3 d and $P>$3 d.  The disk fractions are for
Orion 0.1$\pm$0.03 and 0.2$\pm$0.1 respectively, and for NGC 2264, both
fractions are 0.3$\pm$0.1.

To enable more direct comparison with the RHM results for the inner ONC,
we have converted the Orion FF and NGC 2264 periods to $v$.  The $P$ data
provide an estimate of $v$ rather than $v \sin i$ and are systematically
higher for this reason.  For our purposes, it only matters that the data
within each group are internally consistent.  We have subdivided the $v
\sin i$ and $v$ measurements following RHM according to \teff\ and $L$,
corresponding to a subdivsion by mass ($M$) for stars on convective
tracks, and with age for stars of a given $M$.  For each bin we show in
Table 1 the average $v$ or $v \sin i$.  To first order, each column of the
Table can be viewed as following the evolution of stars of a given mass
down a convective track.  

As can be seen from Table 1, the average value of $v$ or $v \sin i$ {\it
decreases} with $L$ at fixed $M$, rather than increasing as one
would expect for conservation of stellar $J$.  In constructing these
averages for the ONC sample, we have assigned 11 km s$^{-1}$ as the $v
\sin i$ for all of those stars for which only upper limits are available. 
A detailed examination of the data in RHM shows that the percentage of 
stars rotating at or below this upper limit increases with decreasing $L$,
and so the decline in $v \sin i$ with decreasing $L$ is probably even
stronger.

Table 2 tabulates the quantities $v/R$, which is proportional to $\omega$,
and $vR$, which is proportional to $J$.  $v/R$ is constant within each
mass group, within about a factor of $\sim$2, regardless of how far the
stars have evolved down their convective tracks.  In other words, it
appears as if some agent is keeping stars close to constant $\omega$ over
nominal ages from 0.1-3 Myr.  We note that this result stands whether
stars appear to have $I-K$ excesses or not.

\section{Discussion}

Prior to this survey, we expected that PMS stars lacking observable disks
should spin up as they contract along convective tracks.  The observations
show  that they do not, despite the fact that these stars span a range of
0.4 dex in log $R$, which should correspond to a decrease in log $P$ of
0.8 dex.  We briefly explore several possible explanations.

\subsection{Disk-Locking with Gaseous Disks}

The absence of evolution in the distribution of periods among PMS stars as
they contract is consistent with what would be expected for disk-locking.
Absent observable evidence of disks from infrared excess emission arising
from micron-size dust grains, we would need to posit linkage between
stellar magnetospheres and disks in which small solid particles have been
agglomerated into larger bodies, leaving behind only gas.  Detection of
such disks via H$_2$ or CO emission awaits more sensitive measurements
than possible at present. A search for dust-free gaseous disks with SIRTF
appears to be the  most promising near-term approach.

\subsection{Radius Changes Reflect Initial Conditions (Birthline Effects)
Rather Than Evolution}

Suppose that (1) stars in Orion and NGC 2264 were born in a single burst
of star formation $\sim$1 Myr ago; (2) the range in $L$ (equivalently $R$)
for the PMS stars in these clusters does not reflect evolution down
convective tracks, but rather differences in mass accretion rate (\mdot)
that force stars to evolve along different `birthlines' and therefore to
arrive at different initial luminosities along the convective tracks
(Palla \& Stahler 1992); and (3) PMS stars are locked to a particular $P$
so long as they are surrounded by accretion disks. In this picture, the
observed distribution of stars along a convective track for a given mass
then reflects a range of \mdot: protostars with higher \mdot\ have larger
initial radii and lie higher in the color-magnitude diagram following the
end of the accretion phase (see also Hartmann \etal\ 1997). The similarity
of $P$ at different $R$ then follows from the assumption that stars are
locked to their disks for much (nearly all) of their $\sim$1 Myr accretion
phase.

The difficulty with this picture lies with the additional requirement that
stellar $L$ cannot have decreased much since the stars were deposited on
their convective tracks.  If stars that formed through high \mdot\ and
were initially deposited high on their convective tracks had subsequently
evolved downwards, we would expect to see a `tail' of stars that have spun
up to shorter $P$ mixed in with the ensemble of objects that started their
evolution at smaller $R$.  We do not.  The only way to resolve this
contradiction and still maintain conservation of stellar $J$ would be to
identify a mechanism for halting or slowing evolution toward smaller radii
for a time comparable to $\sim$1 Myr.  The problem is even more severe 
when considering those stars in NGC 2264 with apparent ages $\sim$3 Myr, thus
requiring $L$ evolution for stars with high \mdot\ to be delayed for an
even longer time.  

\subsection{Angular Momentum Loss via Stellar Winds}

Stellar winds loaded onto open magnetic field lines can exert a spindown
torque on stars.   However, this mechanism has been shown to be
ineffective for PMS stars because, during this phase of evolution, the
timescale for spindown exceeds the evolutionary timescale by a few orders
of magnitude (\eg\ MacGregor and Charbonneau 1994) --  fully convective
stars are assumed to rotate as solid bodies, and the wind must slow down
the entire star.  Furthermore, this conclusion is nearly independent of
the rate of mass loss.  Calculations (\eg\ Kawaler 1988) show that the
rate of change of $J$ depends on the product of the mass loss rate and the
square of the Alfven radius, but the Alfven radius varies inversely as
some power of the mass loss rate, with the specific power depending on the
configuration of the magnetic field.  For a field geometry ``intermediate"
between a dipolar and a radial field, $dJ/dt$ does  not depend on mass
loss rate at all (Bouvier \etal\ 1997).  The only circumstance under
which  magnetic winds could play an important role in slowing the rotation
of PMS stars would be if there were some way in which to decouple the
outer layers of the star from the interior, \ie\ if $J$ were conserved
locally, and the wind had to slow down only a thin outer layer of the
star.  Such decoupling is not expected for fully convective stars.

Moreover, any putative wind-driven $J$ loss mechanism would have to cease
on timescales of no more than a few Myr in order to account for
the significant population of rapidly rotating main sequence stars in
young clusters such as Alpha Persei (\eg\ Stauffer \etal\ 1989); such
stars require spinup from the PMS to the ZAMS.

\subsection{Angular Momentum Loss via Tidally-Locked Planetary Companions}

Recent theoretical models indicate that tidal locking between a close-in
Jupiter-mass planet and the parent star can transfer spin $J$ from the
star to the orbital $J$ of the planet, thus slowing the rotation of the
star while driving the  planet into a larger orbit. 

Unfortunately, this effect is not large enough to account for the
observations reported  here.  The difficulty lies in the apparent
incompatibility of two requirements: (1) the putative planet must be
located close enough to the star to produce significant tidal distortion;
and (2) the planet must be located far enough from the star to dominate
the $J$ of the system (and thus be able to create significant changes in
stellar $J$ with only modest orbital evolution).  A simple calculation
(\cf\ Trilling \etal\ 1998) in which we assume a Jupiter-mass planet
orbiting a PMS star at a distance of 0.1 AU suggests that $J$ regulation
by tidal locking to such a planet fails by 2 orders of magnitude.

\section{Summary}

Previous attempts to test the disk-locking hypothesis have attempted to
establish a correlation between rotation rates and the presence or absence
of a disk.  We see only weak evidence for this correlation, which has also
been questioned by many previous authors (\eg\ Rebull 2001, Stassun \etal\
1999).  It is possible that intrinsic variations in rotation rates at the
time that stars are deposited on their birthlines is sufficient to mask
such a simple correlation.  What we do see is that the trends in $<v \sin
i>$ and in the $P$ of PMS stars in Orion and in NGC 2264 are consistent
with evolution down convective tracks at constant angular velocity
($\omega$), within a factor of $\sim$2.  The data are not at all consistent
with conservation of stellar angular momentum ($J$). It appears highly
unlikely that the errors in the observations could be large enough to mask
conservation of stellar $J$. The errors in $v \sin i$ and $P$  are small,
and the conclusion that stars do not spin up as they age depends only on
the assumption that the stars are properly ordered according to \teff\ and
$L$; highly accurate calibrations of \teff\ and $L$ are not required.
Conservation of $\omega$ is consistent with the hypothesis that $J$ is
controlled by disk-locking.  However, $\omega$ appears to be conserved, or
nearly so, for both those stars with disks and those for which disks are,
at least so far, undetectable by observation.  We have found no plausible
explanation for this paradox.

We find no evidence for spin-up of stars over the age range spanned by the
current observations, which is $\sim$3 Myr.  Such spin-up must ultimately
occur in order to account for the rapid rotators seen on the main sequence
in young clusters such as Alpha Persei (Stauffer \etal\ 1989).  However,
recent observations of $v \sin i$ for a small sample of stars in the 10
Myr old TW Hya association (Torres \etal\ 2000; Sterzik \etal\ 1999)
suggest mean $v \sin i$ values similar to those found among the oldest
stars in our NGC 2264 sample, despite the fact that the nominal radii
among the TW Hya stars are smaller by 50\%. These observations suggest the
urgent need for a campaign focused on mapping the angular momentum
evolution of low mass stars in the age range 3$-$30 Myr so as to
understand when rotational spinup in response to contraction takes place.

\begin{acknowledgements}
We wish to thank Jonathan Lunine and Lee Hartmann for several helpful
discussions regarding possible explanations for the apparent paradox
explored here. SES acknowledges support from the NASA Origins of Solar
Systems program which enabled analysis of the NGC 2264 data.  We
thank Mark Adams for multiple comments and support during the early phases
of the investigation of periodic stars and the McDonald Observatory for
the award of guest investigator time on the 0.9m telescope. 
\end{acknowledgements}

\begin{deluxetable}{cccccccccccc}
\tablecolumns{12}
\tablewidth{0pc}
\tablecaption{Distribution of $v$ or $v \sin i$.}
\tablehead{
\colhead{cluster} & 
\colhead{log \teff\ $\rightarrow$ bin\tablenotemark{a}} &
\multicolumn{2}{c}{A (3.730-3.682)} &
\multicolumn{2}{c}{B (3.682-3.634)} &
\multicolumn{2}{c}{C (3.634-3.586)} &
\multicolumn{2}{c}{D (3.586-3.538)} &
\multicolumn{2}{c}{E (3.538-3.490)} \\
\colhead{} & 
\colhead{log $L/L_{\sun}$$\downarrow$ bin\tablenotemark{a}} &
\colhead{$<$$v$$>$\tablenotemark{b}} & \colhead{num\tablenotemark{c} } & 
\colhead{$<$$v$$>$} & \colhead{num } & 
\colhead{$<$$v$$>$} & \colhead{num } & 
\colhead{$<$$v$$>$} & \colhead{num } & 
\colhead{$<$$v$$>$} & \colhead{num } } 
\startdata
ONC     & 1 (1.48 - 1.11)    & 33$\pm$13& 3& 41$\pm$10& 5 \\
	& 2 (1.11 - 0.74)    & 30$\pm$4 & 6& 26$\pm$6 & 4 & 14 & 1\\
	& 3 (0.74 - 0.37)    & 26$\pm$5 & 7& 15$\pm$4 & 4 & 34$\pm$12& 7  & 35$\pm$9& 5 \\
	& 4 (0.37 - 0.00)    & 19$\pm$7 & 2& 17$\pm$3 & 13& 17$\pm$3 & 18 & 23$\pm$4& 17 & 34$\pm$15& 3 \\
	& 5 (0.00 - $-$0.37)  & 12       & 1&	     &   & 20$\pm$8 & 6  & 19$\pm$3& 41 & 19$\pm$2 & 22\\
	& 6 ($-$0.37 - $-$0.74)&          &  &          &   & 18$\pm$7 & 3  & 16$\pm$4& 18 & 2-$\pm$3 & 44\\
\hline
Orion FF& 1 &  \\
	& 2 &  380      & 1&          &   & 266      & 1  \\
	& 3 & 34$\pm$7  & 3& 65$\pm$22& 3 & 52$\pm$17& 7  &20$\pm$12 & 2  & 67 & 1\\
	& 4 & 62$\pm$42 & 4& 41$\pm$10& 12& 24$\pm$4 & 17 &58$\pm$13 & 7  & \\
	& 5 &		&  & 12	      &  1& 24$\pm$6 & 18 &37$\pm$8  & 20 & 48$\pm$14& 17\\
	& 6 &		&  & 	      &   & 13$\pm$7 & 2  &32$\pm$7  & 22 & 34$\pm$6 & 35\\
\hline
NGC 2264& 1 &  \\
	& 2 &          &  &          &  & 85 & 1\\
	& 3 & 35       &1 & 59$\pm$33& 2& 55 & 1\\
	& 4 & 21$\pm$8 &6 & 28$\pm$12& 5&36$\pm$20& 3&30$\pm$7  &  2 &  \\
	& 5 & 28       &1 & 23$\pm$11& 5&23$\pm$8 &11&40$\pm$13 & 11 & 36 & 1 \\
	& 6 & 4        &1 & 15       & 1&48       & 1&23$\pm$6  & 18 & 101$\pm$39 & 4\\
\enddata
\tablenotetext{a}{Defined exactly as in RHM; subdivisions according to 
\teff\ and $L$ correspond to subdivsions by $M$ for stars on convective 
tracks, and with age for stars of a given $M$.}
\tablenotetext{b}{$<v \sin i>$ for ONC, $<v>$ for Orion FF and NGC 2264.}
\tablenotetext{c}{Number of stars in the bin.}
\end{deluxetable}

\begin{deluxetable}{cccccccccccc}
\tablecolumns{12}
\tablewidth{0pc}
\tablecaption{Trends in angular momentum ($\propto vR$) and angular 
velocity ($\propto v/R$).}
\tablehead{
\colhead{cluster} & 
\colhead{log \teff\ $\rightarrow$ bin\tablenotemark{a}} &
\multicolumn{2}{c}{A (3.730-3.682)} &
\multicolumn{2}{c}{B (3.682-3.634)} &
\multicolumn{2}{c}{C (3.634-3.586)} &
\multicolumn{2}{c}{D (3.586-3.538)} &
\multicolumn{2}{c}{E (3.538-3.490)} \\
\colhead{} & 
\colhead{log $L/L_{\sun}$$\downarrow$ bin\tablenotemark{a}} &
\colhead{$v/R$\tablenotemark{b}} & \colhead{$vR$\tablenotemark{b}} &
\colhead{$v/R$} & \colhead{$vR$} &
\colhead{$v/R$} & \colhead{$vR$} &
\colhead{$v/R$} & \colhead{$vR$} &
\colhead{$v/R$} & \colhead{$vR$} }
\startdata
ONC     & 1 (1.48 - 1.11)    & 6  &191 &  6 &296 &	  \\
	& 2 (1.11 - 0.74)    & 8  &112 &  6 &123 & 2  &  82  \\
	& 3 (0.74 - 0.37)    & 11 &64  &  5 &46  & 9  &  131 & 7 & 168\\
	& 4 (0.37 - 0.00)    & 12 &31  &  8 &34  & 7  &  43  & 7 & 72 & 9 & 133 \\
	& 5 (0.00 - $-$0.37)  & 11 &13  &    &	& 12 &  33  & 9 & 39 & 7 & 48  \\
	& 6 ($-$0.37 - $-$0.74)&    &    &    &	 & 17 &  19  & 12& 21 & 12& 33  \\
\hline
ONC FF  & 1 &  \\
	& 2 & 116& 1246&   &    & 50 &1428		      \\
	& 3 & 16 & 75  & 22& 197& 16 &177 & 5  & 81 & 12& 375 \\
	& 4 & 32 & 120 & 19& 90 & 10 &63  & 21 & 165	      \\
	& 5 &	 &     & 12& 14 & 14 &44  & 19 & 77 & 21& 116 \\
	& 6 &	 &     &   &    & 11 &16  & 21 & 49 & 21& 57  \\
\hline
NGC 2264& 1 &  \\
	& 2 &	 &    &    &	& 17 & 428		      \\
	& 3 & 13 & 95 & 19 & 191& 15 & 199		      \\
	& 4 & 13 & 38 & 12 & 65 & 15 & 90 & 11 & 87 &	      \\
	& 5 & 23 & 35 & 16 & 35 & 15 & 38 & 20 & 83 & 17&  78 \\
	& 6 & 5  & 4  & 22 & 12 & 35 & 69 & 15 & 38 & 60& 173 \\
\enddata
\tablenotetext{a}{Defined exactly as in RHM; subdivisions according to 
\teff\ and $L$ correspond to subdivsions by $M$ for stars on convective 
tracks, and with age for stars of a given $M$.}
\tablenotetext{b}{$<v \sin i>$ for ONC, $<v>$ for Orion FF and NGC 2264.  
$vR$ would be constant if stellar $J$ is conserved, and $v/R$ would be
constant if $\omega$ conserved.  Note that $vR$ changes, whereas
$v/R$ is constant to within a factor of $\sim$2.}
\end{deluxetable}

\begin{figure*}[tbp]
\epsscale{0.5}
\plotone{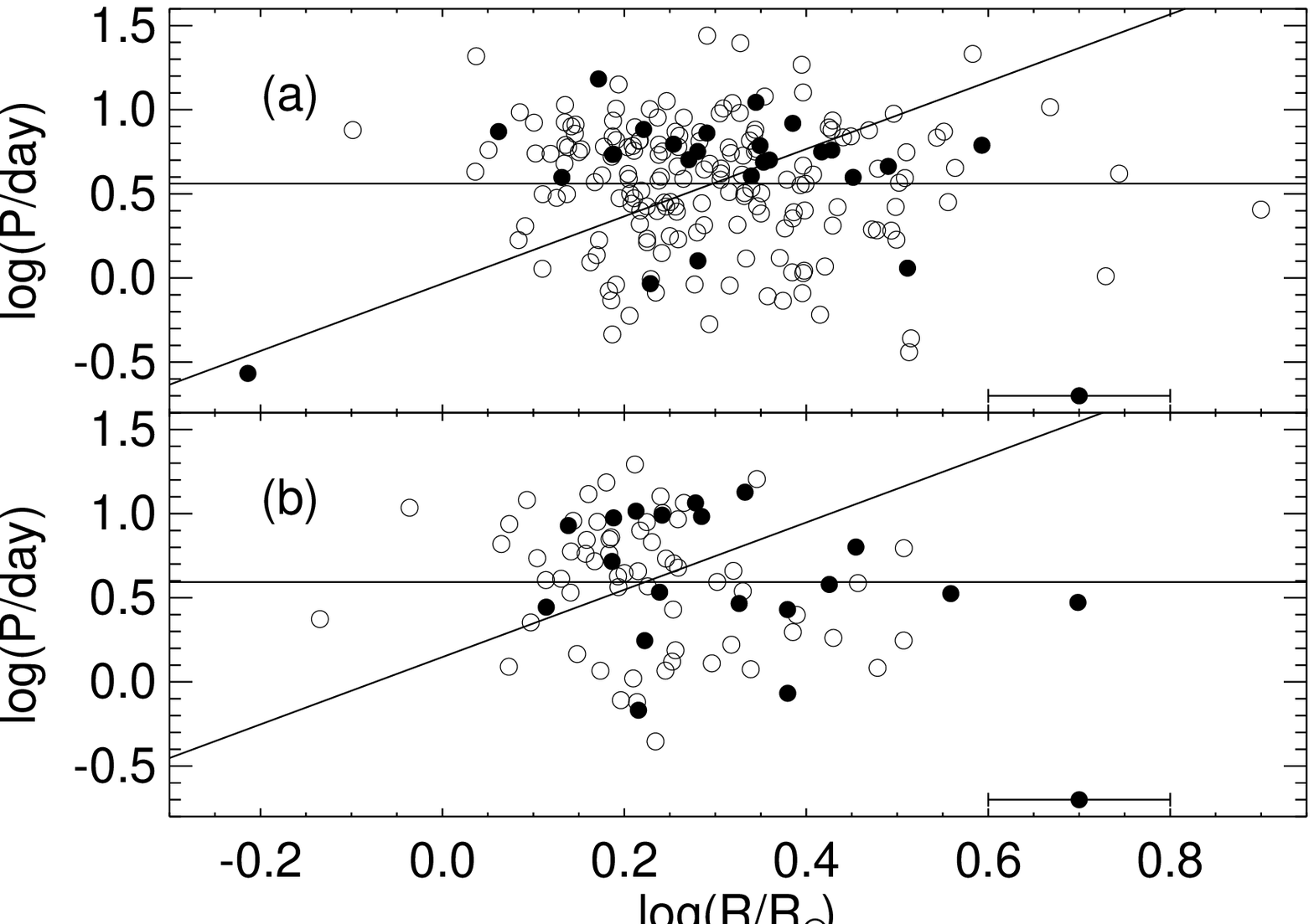}
\plotone{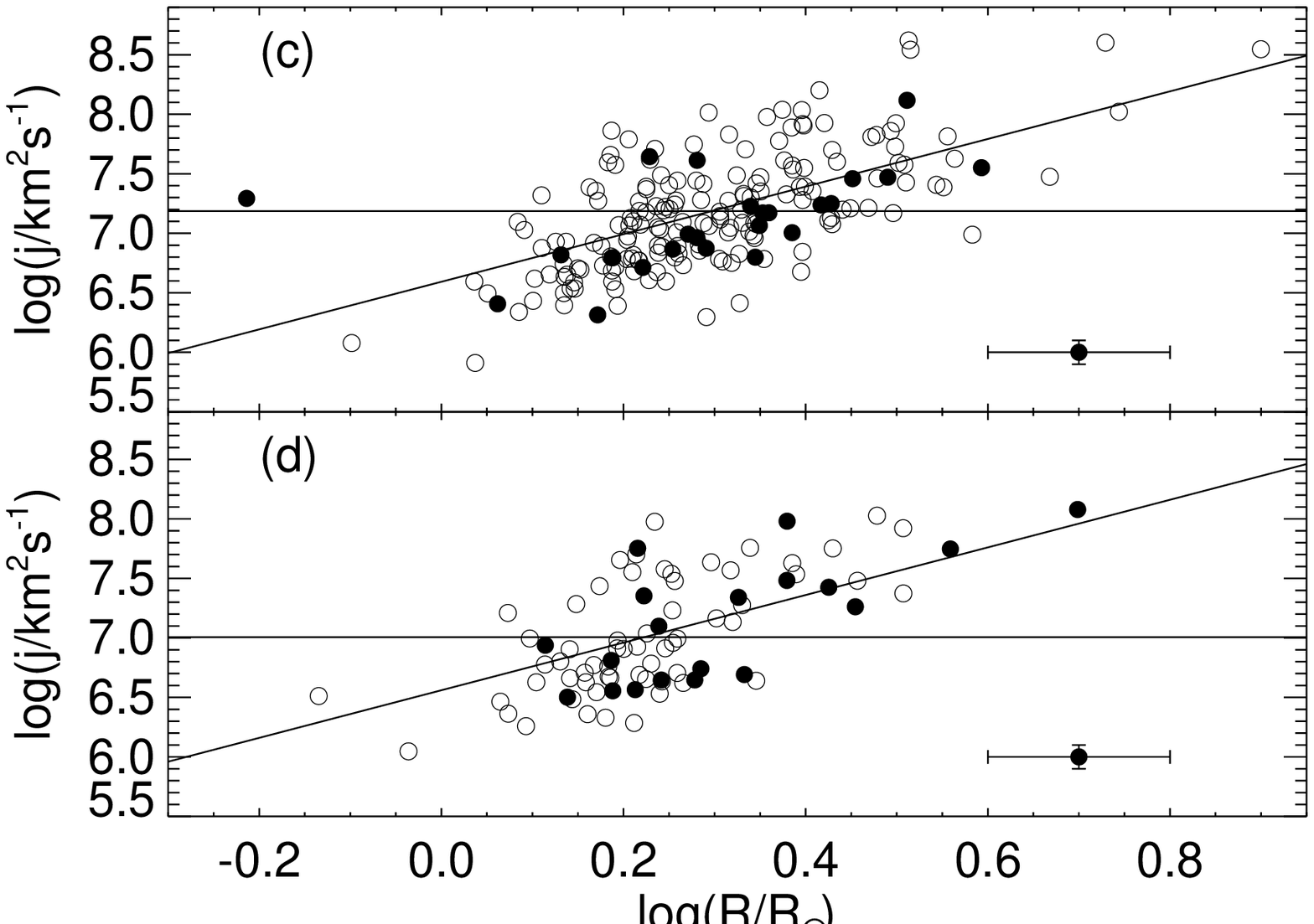}
\caption{Period vs.\ $R$ (a-b) and angular momentum ($j=\frac{2Rv}{5}$)
vs.\ $R$ (c-d) for stars in the Orion Flanking Fields  (a,c) and
NGC~2264 (b,d), without $I-K$ excesses (open symbols) and with excesses
(filled symbols).  Typical error bars are indicated.  In a \& b, lines
correspond to a slope of 0 (constant $P$ or $\omega$ and $j$ changes), and
a slope of 2 ($P$ changes and $j$ conserved); in c \& d, lines also
have slopes of 0 and 2, but slope of 0 corresponds to constant $j$ and
slope of 2 corresponds to constant $P$. }
\end{figure*}

\end{document}